\documentclass[amssymb,twocolumn,aps]{revtex4}
\usepackage{graphics,color,graphicx,amsmath}
\usepackage{subcaption}
\usepackage{natbib}
\usepackage{verbatim}
\usepackage{graphicx}
\usepackage[above,below]{placeins}
\usepackage{float}
\usepackage{tabularx}
\usepackage{times}
\usepackage{blindtext}
\usepackage{url}
\usepackage{rotating}

\renewenvironment{quote}
  {\list{}{\rightmargin=.1cm \leftmargin=.3cm}%
   \item\relax}
  {\endlist}

\begin{document}

\title{Understanding interaction network formation across instructional contexts in remote physics courses}

\author{Meagan Sundstrom,$^1$ Andy Schang,$^1$ Ashley B. Heim,$^2$ and N. G. Holmes$^1$}
\affiliation{$^1$Laboratory of Atomic and Solid State Physics, Cornell University, Ithaca, New York 14853, USA\\
$^2$Department of Ecology and Evolutionary Biology, Cornell University, Ithaca, New York 14853, USA}

\date{\today}

\begin{abstract}
    Engaging in interactions with peers is important for student learning. Many studies have quantified patterns of student interactions in in-person physics courses using social network analysis, finding different network structures between instructional contexts (lecture and lab) and styles (active and traditional). Such studies also find inconsistent results as to whether and how student-level variables (e.g., grades and demographics) relate to the formation of interaction networks. In this cross-sectional research study, we investigate these relationships further by examining lecture and lab interaction networks in four different remote physics courses spanning various instructional styles and student populations. We apply statistical methods from social network analysis -- exponential random graph models -- to measure the relationship between network formation and multiple variables: students' discussion and lab section enrollment, final course grades, gender, and race/ethnicity. Similar to previous studies of in-person courses, we find that remote lecture interaction networks contain large clusters connecting many students, while remote lab interaction networks contain smaller clusters of a few students. Our statistical analysis suggests that these distinct network structures arise from a combination of both instruction-level and student-level variables, including the learning goals of each instructional context, whether assignments are completed in groups or individually, and the distribution of gender and major of students enrolled in a course. We further discuss how these and other variables help to understand the formation of interaction networks in both remote and in-person physics courses.
\end{abstract}

\maketitle

\section{Introduction}
\vspace{-0.5pc}

Social interactions with others, including instructors and peers, are central to student learning~\cite{vygotsky1978,rogoff1996,burkholder2020factors}. Specifically, collaboration with others allows for sharing information and co-constructing understanding while also affording opportunities for reflection and troubleshooting~\cite{olitsky2007promoting,wuchty2007increasing,park2017chemical,chi2014icap}. Students also often feel a stronger sense of belonging and community in a classroom environment when they participate in shared learning experiences~\cite{tinto1997classrooms,irving2020communities}. Within undergraduate science courses in particular, engagement in interactions with peers has been linked to increases in students' self-efficacy, sense of belonging, self-confidence, identity development, and academic achievement~\cite{ballen2017enhancing,sharma2005improving,bjorklund2020connections,dou2016beyond,williams2015understanding,dokuka2020academic,bruun2013talking,traxler2018networks,irving2020communities}. Understanding how students interact with one another, therefore, is important for instructional design.

Many researchers have performed quantitative analyses of student interactions in in-person physics courses~\cite{commeford2020characterizing,commeford2021characterizing,traxler2020network,intergroupinreview,wu2022,brewe2010changing,grunspan2014,zwolak2017students,zwolak2018educational,dou2016beyond,williams2015understanding,traxler2018networks,williams2019linking,brewe2012investigating,bruun2013talking,wells2019}. These studies predominantly use social network analysis, a methodology for visualizing and quantitatively analyzing social structures, to characterize patterns of such interactions. A handful of studies suggest that the structure of interaction networks varies between different instructional contexts (lecture and laboratory) and styles (active and traditional)~\cite{commeford2021characterizing,traxler2020network,intergroupinreview,wu2022,brewe2010changing}. Other studies, moreover, find discrepant results pertaining to how students' positions in an interaction network relate to their attributes (e.g., grades and demographics)~\cite{dokuka2020academic,dou2016beyond,williams2015understanding,wells2019,brewe2012investigating}. 

To reconcile these inconsistent results about interaction network formation, we investigate student interactions in four different remote physics courses spanning different instructional contexts and styles, student populations, and semesters. We apply statistical methods of social network analysis to examine how students' laboratory (lab) and discussion section enrollment, final course grades, gender, and race/ethnicity relate to network formation within these various instructional conditions.

\vspace{-0.75pc}
\subsection{What variables relate to social network formation in physics courses?}
\vspace{-0.5pc}

Previous studies of in-person courses suggest that both instruction-level variables, such as context (lecture or lab) and style (active or traditional), and student-level variables, such as grades, gender, and race/ethnicity, relate to interaction network formation. 

\vspace{-0.75pc}
\subsubsection{Instruction-level variables}
\vspace{-0.5pc}

Most quantitative studies of interactions investigate networks at the course level. These studies use surveys asking students to report interactions they have had with peers about anything in their physics course~\cite{brewe2010changing,traxler2020network,grunspan2014,zwolak2017students,zwolak2018educational,dou2016beyond,williams2015understanding,traxler2018networks,williams2019linking,brewe2012investigating,wells2019}. Because introductory physics courses contain multiple instructional contexts (e.g., lecture, discussion sections, labs), these ``course-wide'' surveys likely capture interactions about a variety of course material. Lectures and discussion sections (where students work in small groups on physics problems related to lecture content), however, often have different learning objectives and cover different material than labs~\cite{Phys21,kozminski2014aapt,holmes2018introductory,Smith2021}. Correspondingly, one study found that course-wide interaction networks have different structures than lab-specific interaction networks, where students only report interactions with their lab peers~\cite{commeford2021characterizing}. They observed that the course-wide interaction networks contain large clusters of connections between many students, while lab interaction networks contain smaller, isolated clusters of a few students (likely representing defined lab groups). This result suggests that the structure of interaction networks likely varies across instructional context, whether lecture or lab.


One reason for this distinction might be the course structure itself. Students gain exposure to different peers in each course component (e.g., lecture, discussion sections, labs) and frequently interact with peers in their discussion and lab sections through small group work~\cite{commeford2021characterizing,dokuka2020academic}. Students may discuss different course material with these discussion and lab peers, however, which might explain the different structures of lab and lecture interaction networks. In the current study, therefore, we ask students to report interactions about lecture and lab material separately. We also quantitatively examine the relationship between students' discussion and lab section enrollment and network formation.


Instructional style is also related to interaction network formation. Researchers have compared the networks of courses implementing active learning (student-centered teaching that promotes interactive engagement) and traditional instruction (transferring information from instructor to students through lectures)~\cite{brewe2010changing,intergroupinreview,wu2022}. Brewe and colleagues~\cite{brewe2010changing}, for example, report on the interaction networks of two introductory physics courses, one using an active learning pedagogy with lots of group work and one using traditional lectures. They found a significant increase in network connectedness (the proportion of observed to possible network connections among students) between the beginning and the end of the semester in the active learning course, but not in the traditional course. The end-of-semester networks in each course, moreover, had very different structures. The active learning course network contained long chains of connections among all of the students, whereas the traditional course network contained some small clusters of students and many isolated students remaining completely disconnected from their peers. This study suggests that network structure might vary by instructional style, whether active or traditional. 

Even between different types of active learning instruction, however, different network structures emerge~\cite{commeford2021characterizing,traxler2020network}. For example, Commeford and colleagues~\cite{commeford2021characterizing} examined two in-person active learning physics courses, one with many whole-class discussions and one centering around small group work. The interaction network in the first course was highly connected, while the network in the second course contained many isolated clusters of a few students. This result adds nuance to the relationship between instructional style and interaction networks: coarse-grained categories of instruction, such as active and traditional~\cite{stains2018anatomy,commeford2022latent}, do not fully explain differences in network structure. Instead, the particular instructional techniques implemented in a course may impact students' patterns of interactions. In the current study, therefore, we examine interaction networks across various traditional and active learning courses.

\vspace{-0.75pc}
\subsubsection{Student-level variables}
\vspace{-0.5pc}

In addition to instruction-level variables, research indicates that students' course grades, gender, and race/ethnicity may relate to their patterns of interactions. 


First, many studies have found that students who have more and/or stronger connections to peers earn higher grades. This correlation could either be due to students performing well on an assessment and then engaging in more peer interactions (such as due to increased confidence or peers seeking their help) or students learning from their peer interactions and subsequently performing well on assessments~\cite{williams2019linking}. This correlation has been previously observed between students' positions in course-wide interaction networks and their overall course grades~\cite{williams2019linking, williams2015understanding,dokuka2020academic,grunspan2014,traxler2018networks,stadtfeld2019integration} as well as between students' engagement in lab-specific interactions and their lab grades~\cite{wei2018developing,park2017chemical}. 

Other work, however, suggests that the correlation between interactions and performance varies by the course material being discussed. 
For example, Bruun and Brewe~\cite{bruun2013talking} investigated physics students' interactions about the conceptual and problem solving aspects of the course separately. They found that students with higher numbers of connections to their peers in the conceptual physics interaction network tended to score higher grades in the course. In the problem solving interaction network, however, it is students who are connected to well-connected peers that earn higher grades. This result suggests that within different instructional contexts, different kinds of network positions correlate with students' grades. To further examine this possibility, we separately analyze the relationship between students' positions in lecture and lab interaction networks and their final course grades. 

Second, prior work observed mixed results with regard to whether and how students' gender relates to their network position. Research has found that men hold more central positions than women in the friendship network of a cohort of undergraduate economics majors and the interaction networks of introductory physics students at a large institution~\cite{dokuka2020academic,dou2016beyond,williams2015understanding}. Another study, however, found that women hold more central positions than men in the interaction network of an introductory physics course for physics majors at a small liberal arts college~\cite{wells2019}. Still another study observed that men and women hold equally central positions in the interaction network of an informal physics learning center~\cite{brewe2012investigating}. To contribute to our understanding of how the relationship between gender and network formation varies across courses serving different student populations (e.g., different gender enrollments) and implementing different instructional styles, we examine interaction networks in multiple different physics courses.

Finally, students' race/ethnicity seems to be uncorrelated with their network position~\cite{zwolak2017students,williams2015understanding,brewe2012investigating}. This body of work, however, only analyzed physics courses in which racially/ethnically minoritized students comprise most of the student population. We expand on these studies by investigating the the relationship between students' race/ethnicity and network formation within different student populations and in remote courses.

\vspace{-0.75pc}
\subsection{Interactions in in-person versus remote courses}
\vspace{-0.5pc}

The research summarized so far examined the interaction networks of in-person physics courses, however the COVID-19 global pandemic necessitated remote instruction at many universities. It is possible that the nature of student interactions differs between in-person courses and remote courses. For example, students in in-person courses have easy access to peers and instructors in the same room, whom they can collaborate with or ask for help. In contrast, students in remote courses might work more independently if they do not have adequate internet access or if the instruction does not involve collaboration among peers or offer means of meeting new peers \cite{rosen2020epistemology}.

Researchers have a limited understanding, however, of student interactions during remote instruction and whether or not they align with those in in-person courses. A handful of studies have examined interactions in online labs~\cite{reeves2021virtual,wei2019understanding,rosen2020epistemology,attardi2018improving, rosen2021}. For example, one study found that students in an in-person undergraduate physics lab value socialization more than their peers in an online lab~\cite{rosen2020epistemology}. Other studies directly investigated the impacts of the COVID-19 pandemic on student interactions in their remote courses more generally. These studies~\cite{wilcox2020recommendations,hussein2020exploring,karalis2020teaching,kyne2020covid,gillis2020covid19,dew2021student,rosen2021,doucette2021newtothis,marzoli2021,klein2021studying,conrad2021} unanimously found that undergraduate students engage in fewer interactions with their peers during remote instruction compared to in-person instruction. This body of work, however, mostly uses questionnaires to probe students' perceived experiences of their interactions. In the current study, we aim to expand our understanding of student interactions in remote physics courses by analyzing students' actual reported interactions with social network analysis methods.


\vspace{-0.75pc}
\subsection{Current study}
\vspace{-0.5pc}

Research on in-person physics courses suggests that different interaction network structures emerge in lecture and lab instructional contexts. Other studies offer possible explanations for these varying network formations, such as instructional style and student attributes, however they find inconsistent patterns. We investigate these explanations further by examining student interactions in multiple different remote physics courses. The following research question guided our study: Between the instructional contexts of lecture and lab, how do instruction-level and student-level variables relate to the formation of interaction networks?  

To address this research question, we implemented a cross-sectional research design to observe student interaction networks at a given point in time and measure the relationships between relevant explanatory variables and these networks. Specifically, we administered a network survey in four different remote, introductory physics courses asking students to self-report peers with whom they have had meaningful interactions about lecture and lab material. We then applied statistical methods from social network analysis -- exponential random graph models -- to measure how students' section enrollment, final course grades, gender, and race/ethnicity relate to the formation of the networks. 

We find that, similar to studies of in-person courses, lecture and lab networks exhibit different network structures. No single variable, however, fully accounts for the formation of these networks. Instead, network structure is related to a combination of variables: the learning goals of various instructional contexts, the pervasiveness of different course material, students' grades, whether assignments are completed in groups or individually, the distribution of gender and major of students enrolled in a course, and the tendency for students to interact with peers of their same gender. Notably, students' race/ethnicity seems unrelated to their position in interaction networks in both in-person and remote physics courses.

\vspace{-0.75pc}
\section{Methods}
\vspace{-0.5pc}

In this section, we first describe the four courses analyzed in this study. Then, we provide details about the network survey we administered and outline the statistical methods used for analysis.


\begin{table*}[t]
\caption{\label{tab:demographics}%
Summary of the semester, course, and modality of each course as well as the self-reported gender, URM (underrepresented and minoritized) status, intended major, and academic year of students in each course. Numbers in parentheses are the $N$ values corresponding to the percentages. All online components were held synchronously on Zoom.}
\begin{ruledtabular}
\setlength{\extrarowheight}{1pt}
\begin{tabular}{lcccc}
\textrm{}&
\textrm{M-Eng}&
\textrm{M-Phys}&
\textrm{EM-Eng}&
\textrm{EM-Phys}\\
\colrule
Semester & Fall 2020 & Fall 2020 & Spring 2021 & Spring 2021\\
Course & Mechanics & Mechanics & Electromagnetism & Electromagnetism\\
Modality \\
\hspace{5mm}Lecture sections & 2 Online & 1 Online & 2 Online & 1 Online\\
\hspace{5mm}Discussion sections & 12 Online, 2 In-person & 3 Online, 2 In-person & 8 Online, 4 In-person & 4 Online\\
\hspace{5mm}Lab sections & 14 Online & 5 Online & 12 Online & 4 Online\\
Total enrollment & 208 & 89 & 190 & 56\\
Students in analysis & 198 & 84 & 163 & 43\\
Gender \\
\hspace{5mm}Men & 42\% (84) & 70\% (59) & 33\% (54) & 54\% (23)\\
\hspace{5mm}Women & 47\% (92) & 28\% (23) & 39\% (63) & 23\% (10)\\
\hspace{5mm}Non-binary & 0 & 1\% (1) & 0.6\% (1) & 0\\
\hspace{5mm}Unknown & 11\% (22) & 1\% (1) & 28\% (45) & 23\% (10)\\

 Race/ethnicity\\
\hspace{5mm}Non-URM & 71\% (140) & 81\% (68) & 58\% (95) & 70\% (30)\\
\hspace{5mm}URM & 16\% (32) & 14\% (12) & 12\% (21) & 7\% (3)\\
\hspace{5mm}Unknown & 13\% (26) & 5\% (4) & 29\% (47) & 23\% (10)\\
 
Major\\
\hspace{5mm}Physics/Engineering Physics & 5\% (11) & 69\% (58) & 6\% (10) & 72\% (31)\\
\hspace{5mm}Engineering & 65\% (128) & 17\% (14) & 57\% (93) & 2\% (1) \\
\hspace{5mm}Other (STEM) & 10\% (19) & 8\% (7) & 6\% (10) & 2\% (1) \\
\hspace{5mm}Unknown & 20\% (40) & 6\% (5) & 31\% (50) & 24\% (10) \\

Year \\
\hspace{5mm}First-year & 83\% (166) & 93\% (78) & 56\% (92) & 74\% (32) \\
 \hspace{5mm}Second-year & 12\% (24) & 4\% (3) & 13\% (22) & 2\% (1) \\
\hspace{5mm}Third-year & 3\% (6) & 1\% (1) & 4\% (7) & 0 \\
\hspace{5mm}Other/Unknown & 2\% (2) & 2\% (2) & 26\% (42) & 24\% (10) \\
\end{tabular}
\end{ruledtabular}
\end{table*}

\vspace{-0.75pc}
\subsection{Courses and participants}
\vspace{-0.5pc}

Our study includes two course sequences inclusive of four calculus-based introductory physics courses at Cornell University: two mechanics courses from fall 2020 and two electromagnetism courses from spring 2021. One course sequence is intended for students majoring in engineering or other STEM disciplines, while the second is intended for physics majors. We will refer to the mechanics course for engineering majors as ``M-Eng," the mechanics course for physics majors as ``M-Phys," the electromagnetism course for engineering majors as ``EM-Eng," and the electromagnetism course for physics majors as ``EM-Phys."

Table \ref{tab:demographics} summarizes the four courses by mode of instruction, enrollment, and students' self-reported demographics. All lectures were held synchronously online through Zoom and in all four courses a male faculty member in the physics department instructed the lectures. M-Eng and EM-Eng lectures were taught using active learning techniques: both were flipped courses where students read text or watched pre-lecture videos and completed reading quizzes before attending lecture. Lectures for M-Eng used conceptual poll questions and instructor demonstrations and lectures for EM-Eng used math-based problems through Learning Catalytics~\cite{newland2021review}. In both M-Eng and EM-Eng, students answered questions both individually and following group discussion in Zoom breakout rooms during lectures. M-Phys and EM-Phys lectures followed a traditional instruction style, with the instructor spending most of the time presenting material and working through example problems. In M-Phys lectures, the instructor also used poll questions that students answered individually. EM-Phys did not use poll questions. In all four courses, students completed long, individual homework assignments (problem sets) each week.

Graduate teaching assistants instructed the discussion and lab sections for each course. Discussion sections met twice per week for 50 minutes and lab sections met once per week for two hours. Each discussion and lab section contained approximately 20 students who worked together in small groups of two to four students. Most of the discussion sections and all of the lab sections took place synchronously online through Zoom where students worked in groups in virtual breakout rooms. In the few discussion sections held in person, students worked together at round tables. 

During discussion sections, students worked together to solve problems related to lecture content. In M-Eng, M-Phys, and EM-Phys, discussion problems were completed as a group but students did not submit any work. In EM-Eng, students solved problems through Learning Catalytics~\cite{newland2021review} and submitted their work as a group. The formation of small groups during discussion sections varied by teaching assistant, with some formed randomly and some formed based on student preference (though there were no formally administered surveys probing these preferences). The individual teaching assistants also decided whether discussion groups changed or remained the same throughout the semester (we do not have this information for individual sections). 

Labs in every course were inquiry-based (as per the work described in, for example,~\cite{holmes2018introductory,Smith2021,smith2020direct,holmes2015teaching, kalender2021}) and students designed experiments using objects at home or in their dorm room. Lab groups were formed based on a group-forming survey where students could indicate their preferences related to group gender composition and role division and list the names of peers they did or did not want to work with. Teaching assistants created lab groups using these reported preferences and also avoided groups containing isolated women. These groups were held the same for the whole semester (with minor adjustments if students withdrew from the course or groups were having collaboration challenges). Students submitted lab notes as a group, rather than individually, and these notes were graded by the teaching assistants. Lab groups typically collaborated on an online document for the notes so that all group members could contribute simultaneously. Students also completed short, individual lab homework assignments each week.

All courses contained a majority of first-year students (see Table \ref{tab:demographics}). Similar proportions of men and women were enrolled in the M-Eng and EM-Eng courses, while more men than women were enrolled in the M-Phys and EM-Phys courses. Additionally, all four courses had a majority of non-URM (underrepresented and minoritized) students. We designate non-URM students as those identifying their race/ethnicity solely as White and/or Asian/Asian American and URM students as those identifying as at least one of any other race/ethnicity (including Black or African American, Hispanic/Latinx, and Native Hawaiian or other Pacific Islander, as defined by the American Physical Society~\cite{racecategories}), defined relative to the physics discipline~\cite{degreesbyrace}.

About 55\% of the students in our data set are represented in two of the four analyzed courses (one mechanics course and one electromagnetism course) because students taking mechanics in the fall typically go on to take electromagnetism in the spring. Many students take the two courses within one course sequence, however some students switch course sequences between semesters (e.g., if they found their mechanics course too challenging or too easy). A small fraction of analyzed students, therefore, are represented in M-Eng and EM-Phys (0.8\%) or M-Phys and EM-Eng (6\%). The remaining 45\% of analyzed students took only a mechanics course or only an electromagnetism course during the surveyed semesters. We suspect that the students taking only a mechanics course either needed just one physics course to fulfill their major requirements, delayed taking the electromagnetism course to a future semester, or dropped out of the course sequence. The students taking only an electromagnetism course likely entered the university with transfer or high school credits that covered the mechanics course.

\vspace{-0.75pc}
\subsection{Data collection}
\vspace{-0.5pc}

Prior work has demonstrated that peers develop a community among one another by about halfway through the semester~\cite{williams2019linking}. Therefore, we administered a network survey in each of the four courses around the halfway point of the 15-week semester. Students completed the survey online via Qualtrics as part of a lab homework assignment about their group work experiences. Our two survey prompts adopted the language used in previous studies~\cite{zwolak2018educational,traxler2020network,dou2019practitioner,commeford2021characterizing} and asked students to self-report peers with whom they had meaningful interactions about different instructional material:
\begin{quote}
    Please list any students in this physics class that you had a meaningful interaction$^*$ with about lab material this week.
    \\
    \\
   Please list any students in this physics class that you had a meaningful interaction$^*$ with about other aspects of the course this week.
   \\
   \\
    $^*$A meaningful interaction may mean virtually over Zoom, through remote chat or discussion boards, or any other form of communication, even if you were not the main person speaking or contributing.
\end{quote}
We refer to the first prompt as ``lab'' interactions and the second prompt as ``lecture'' interactions.

The survey was in an open response format (one text box per prompt) and students could respond with an unlimited number of names. This format avoids students feeling obligated to fill a quota and write down extra names of peers with whom they may not perceive themselves as having had a meaningful interaction~\cite{grunspan2014}. Students were also not given a class roster from which to choose or look up names. This resulted in some listings being hard to match to the class roster during analysis, as there were instances of students misspelling peers' names or reporting just a first or a last name. Details about how we processed the text responses to extract all of the reported interactions are in the Appendix. 

Network measures are robust to up to 30\% of missing data (e.g., due to non-responders)~\cite{kossinets2006effects,smith2013structural} and are more robust with denser networks (networks containing many connections). In M-Eng, M-Phys, and EM-Eng, the survey response rate was over 75\%. Our results for these courses, therefore, are well grounded. In EM-Phys, about 60\% of enrolled students completed the survey. Networks for this course (presented in the next section), however, were particularly dense -- students on average reported multiple peers' names. This adds some validity to our analysis for this course because we have information about many edges and the network likely captures some interactions involving non-respondents, but our claims related to this course should still be considered tentative. We chose not to impute any interactions because the interdependent nature of network data means any imputations may substantially change the properties of the network~\cite{dou2019practitioner}.

We included all students who responded to the survey and/or were listed by at least one peer in our analysis. Students who responded to the survey but did not report any interactions and were not listed by any peers are represented as isolates (zero connections). We also only included the interactions reported by students who consented to participate in research. If a consenting student listed a non-consenting student, we included the interaction in our analysis, but removed all information (e.g., demographics) about the non-consenting student. In all courses, more than 75\% of enrolled students are represented in our analyzed data (see Table \ref{tab:demographics}).

At the end of the semester, we also identified in which discussion and lab section students were enrolled and collected students' final course grades.

\vspace{-0.75pc}
\subsection{Data analysis}
\vspace{-0.5pc}

With these student data and survey responses, we turned to methods of social network analysis~\cite{grunspan2014,dou2019practitioner,brewe2018guide}. We visualized the data as eight different networks, four separate courses each with two instructional contexts (lecture and lab). Each network's \textit{nodes} represent students and the undirected \textit{edges} represent a reported interaction between two students regardless of which student reported the interaction. We examined the network diagrams to determine distinguishing structural features across courses and contexts.

Similar to previous studies~\cite{williams2019linking,traxler2020network,wells2019,bruun2013talking,brewe2010changing,williams2015understanding,zwolak2018educational,zwolak2017students,dou2016beyond,stadtfeld2019integration}, we treated the networks as directed for our statistical analysis. In directed networks, the direction of each edge corresponds to which student reported interacting with the other student. A one-way edge indicates that only one student in a pair reported interacting with the other, while a two-way edge indicates that both students in a pair reported interacting with each other. 
Interactions, however, inherently consist of two-way edges because two students must communicate with one another for an interaction to occur. One-way edges, therefore, could be due to recall bias (e.g., the other student forgot about the interaction or did not know the other person's name and so did not report it) or over-reporting (e.g., one student reported an interaction that the other student did not perceive as meaningful)~\cite{grunspan2014}. Thus, converting all one-way edges to two-way edges (i.e., treating the network as undirected) may over-estimate the total number of interactions in the network, while eliminating all one-way edges may under-estimate the total number of interactions. We therefore kept the one- \textit{and} two-way edges in the network to accurately reflect the survey responses. This quantitative treatment also amplifies the difference between students with many connections and students with few connections, reducing possible statistical noise.

We first calculated each directed network's \textit{density} -- the proportion of all possible edges in the network that we observed -- to gain a sense of the overall level of connectedness among students. We determined the standard errors of the densities via bootstrapping: resampling the observed network many times, calculating the density of each sampled network, and then determining the standard deviation of the densities for all of the sampled networks~\cite{traxler2020network,snijders1999non}. The bootstrapping was performed with 10,000 bootstrap trials for each network using the \textit{snowboot} package in R~\cite{chen2019snowboot}. We then focused our analysis on exponential random graph models (ERGMs). 

\begin{figure}[t]
    \centering
    \includegraphics[width=3.3in]{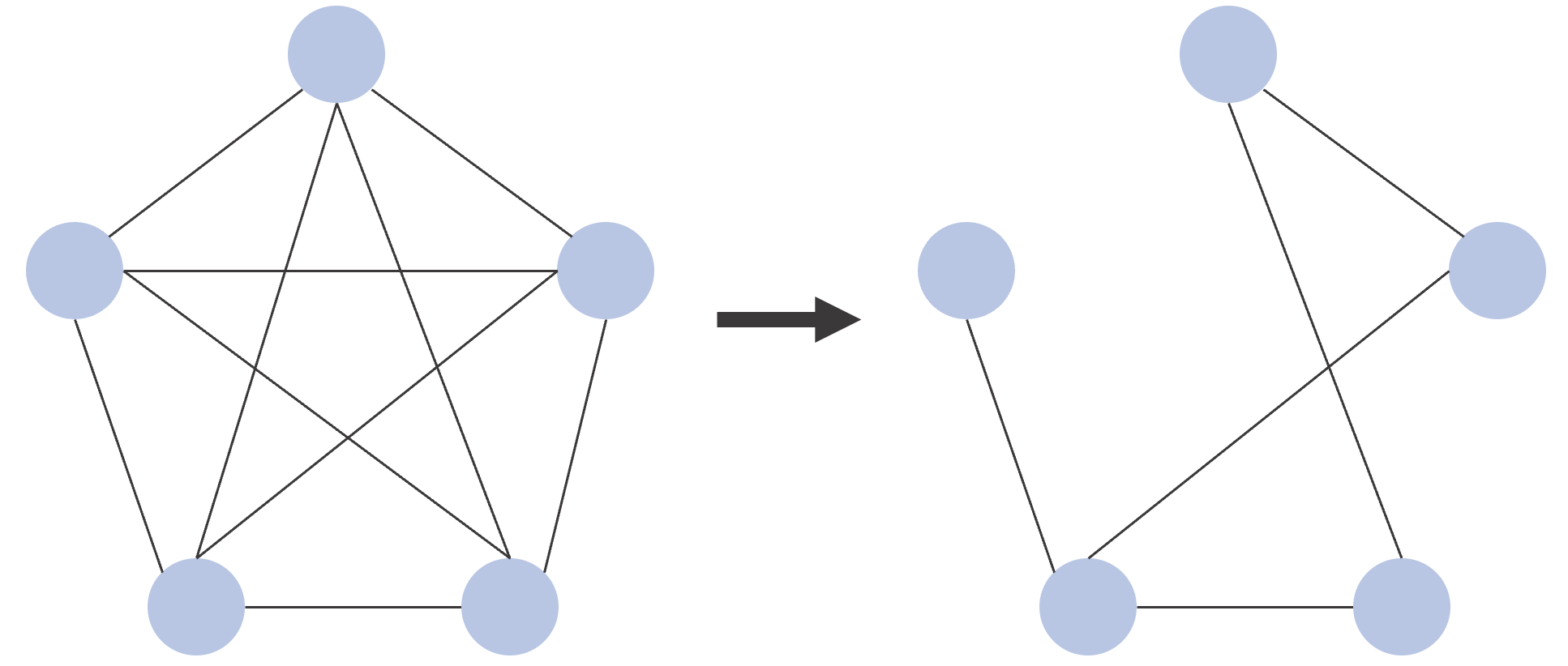}
    \caption{Network of five nodes with all edges present (left) and one possible realization of this network (right). 
    }
    \label{fig:realization}
\end{figure}

ERGMs allow us to determine the important structures or configurations in an observed network~\cite{anderson1999,robins2007,lusher2013exponential}. Such models assume that a network's set of nodes is fixed and that the set of observed edges among the nodes is a realization from a random graph that comes from a distribution belonging to the exponential family. To illustrate this, consider an undirected network containing five total nodes, as shown in  Fig. \ref{fig:realization}. The network on the left shows all possible edges among the five nodes and the network on the right shows one of the many possible realizations or specific instances of the left-hand network that may emerge due to some social process(es). We use ERGMs to infer what process(es) specifically occurred to form this particular realization. For instance, certain social selection processes, such as women interacting more frequently with other women than with men, may have influenced the formation of the realized network. 

Mathematically, we can think of ERGMs as having a similar form to logistic regression models, though the assumption of independence of observations is relaxed. To formulate a model, we choose a principled set of predictor variables (i.e., configurations) that might be related to the formation of the observed network. These variables may be structural (e.g., measuring the tendency for two-way nominations) or nodal (e.g., measuring the extent to which students of a certain gender are more likely to have connections). The goal is to use these $k$ network statistics $g_k(y)$ and their corresponding coefficients $\theta_k$ to predict the structure of the random network $Y$. The model takes the form
\begin{equation*}
    P_\theta[Y = y] = \frac{1}{\psi}\exp\left(\sum_{k} \theta_k g_k(y)\right)
\end{equation*}
where $y$ is a realization of the random network $Y$ and $\psi = \sum_y \exp\left(\sum_{k}\theta_k g_k(y)\right)$ is a normalization constant that ensures that the probability sums to one. Given an observed network $y$, the coefficients of the model are estimated using Maximum Likelihood Estimation (MLE). Due to the dependence between the network edges, the MLE is commonly approximated with Markov Chain Monte Carlo (MCMC) techniques~\cite{hunter2008}. 

There are two different ways to interpret the coefficients of ERGMs. In general, the coefficients weight the importance of each modeled configuration for the formation of the realized network, where positive (negative) coefficients show that the configuration is observed more (less) frequently than by chance after accounting for all other configurations that are modeled. The second way to interpret the coefficients is to focus on specific edges of the network. In this interpretation, the coefficient $\theta_k$ of the $k$th configuration shows how the log-odds of an edge being present changes if the formation of the edge increases the $k$th configuration by one unit, holding the rest of the network constant. For instance, if the predictor variable measures the number of two-way edges in the network, its coefficient represents how much the log-odds of an edge being present increases when the addition of this edge would reciprocate an existing edge.

In our analysis, we included both structural and nodal predictor variables in the model. For the nodal variables, we focused on those related to degree (the number of adjacent edges connected to a node), rather than other measures of centrality (such as closeness and betweenness), because degree has been shown to be relevant in describing network structure and to consistently predict learning outcomes~\cite{saqr2022curious,traxler2022networks}. Therefore, we initially considered three separate ERGM models with the nodal predictor variables depending on (i) \textit{indegree} (number of other students who reported interacting with a given student), (ii) \textit{outdegree} (number of students with whom a given student reported interacting), or (iii) \textit{total degree} (sum of indegree and outdegree). Results from the models using indegree and outdegree variables, however, were not much different than and did not add any nuance to those from the total degree model. In some instances, the indegree and outdegree models also did not fit the observed network sufficiently well. For these reasons, we decided to use total degree (hereon referred to as ``degree") for all nodal predictor variables. 

Our final model included the ten predictor variables listed below. The first three variables provided information about the structure of each network, while the remaining variables measured student-level attributes that might be related to the formation of these networks. For example, the fourth and fifth variables relate students' section enrollment to the network structure. The sixth variable relates students' network positions to their final course grades. Finally, the last four variables compare network positions across demographic groups:




\begin{figure*}[t]
    \centering
    \includegraphics[width=6.2in]{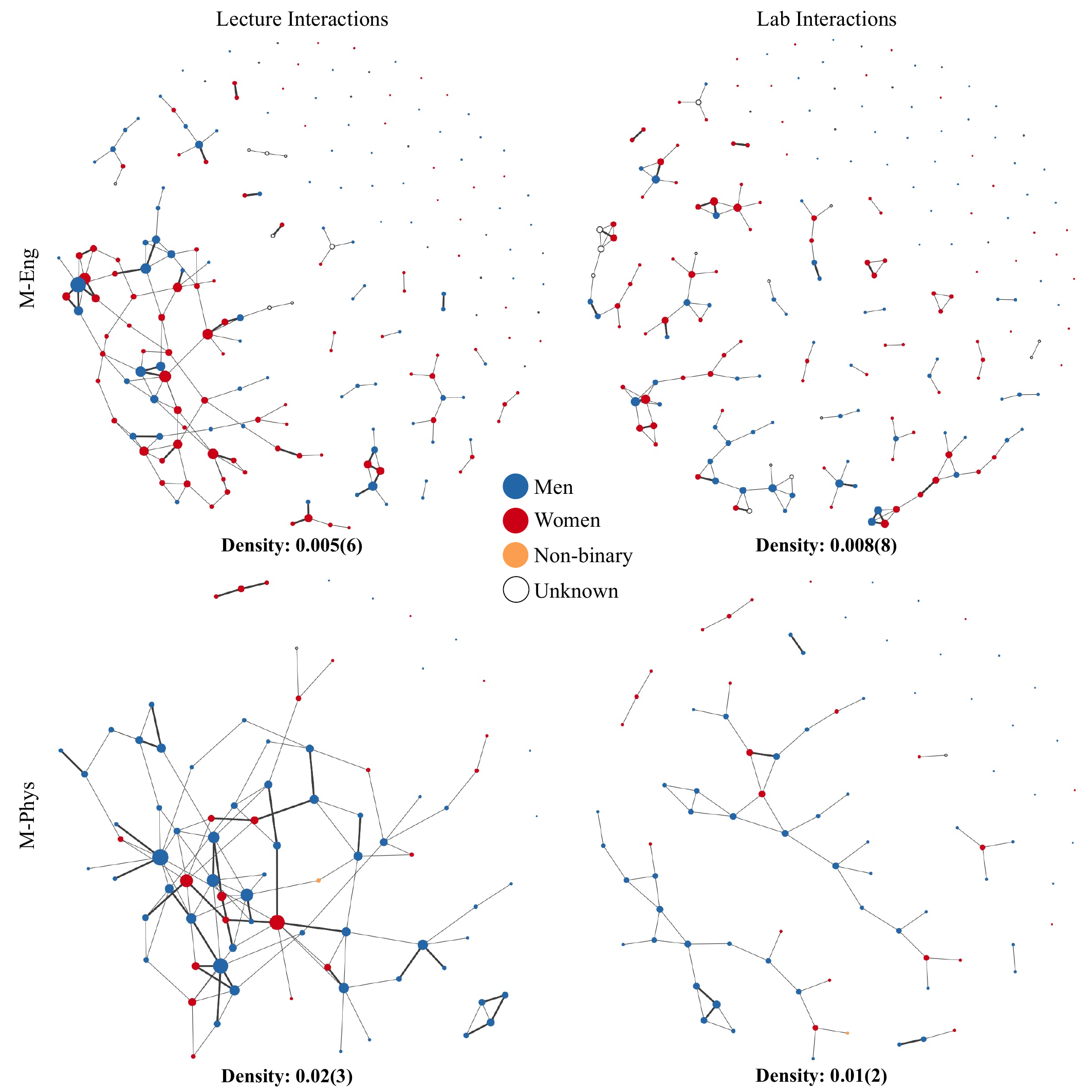}
    \caption{Diagrams and densities of interaction networks for M-Eng and M-Phys. Nodes are colored by gender and sized proportional to total degree (number of adjacent edges). Thick edges represent reciprocal edges (students A and B both reported interacting with one another) and thin edges represent one-way edges (student A reported interacting with student B, but student B did not report interacting with student A). Densities are the proportion of observed to possible edges, with standard errors of the last digit shown in parentheses. These same network diagrams with nodes colored by race/ethnicity are in the Supplementary Material.}
    \label{fig:fallsociograms}
\end{figure*}

\begin{figure*}[t]
    \centering
    \includegraphics[width=6.2in]{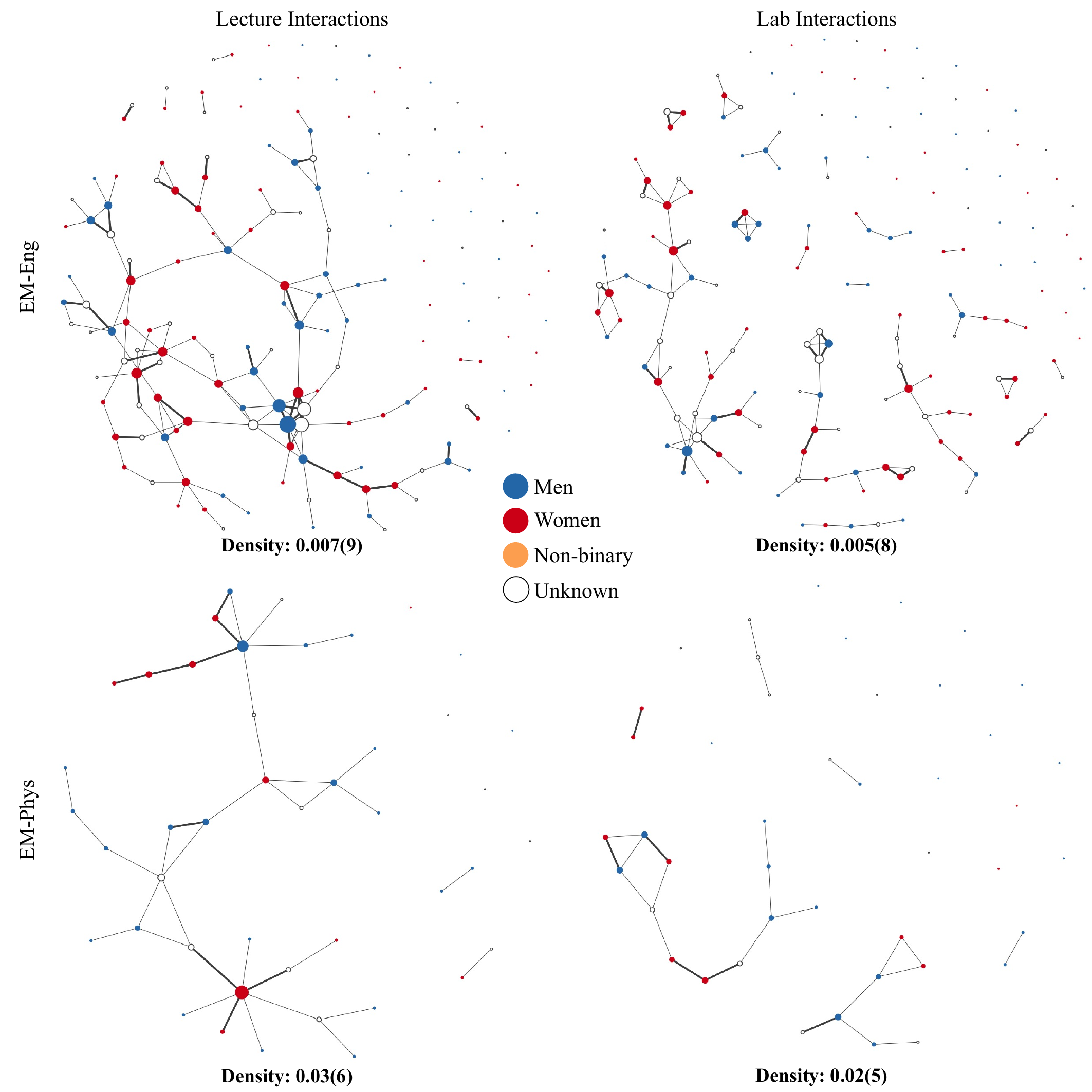}
    \caption{Diagrams and densities of interaction networks for EM-Eng and EM-Phys. Nodes are colored by gender and sized proportional to total degree (number of adjacent edges). Thick edges represent reciprocal edges (students A and B both reported interacting with one another) and thin edges represent one-way edges (student A reported interacting with student B, but student B did not report interacting with student A). Densities are the proportion of observed to possible edges, with standard errors of the last digit shown in parentheses. These same network diagrams with nodes colored by race/ethnicity are in the Supplementary Material.}
    \label{fig:springsociograms}
\end{figure*}

\begin{enumerate}
    \itemsep0em 
    \item \textit{Edges}: main intercept term measuring the number of observed edges
    \item \textit{Reciprocity}: measure of reciprocal edges (e.g., student A reports an interaction with student B and student B reports an interaction with student A)
    \item \textit{Geometrically-weighted edgewise shared partners (GWESP)}; decay parameter = 0.25 as commonly used in the ERGM literature ~\cite{hummel2012improving,yin2021highly,butts2014introduction}): measure of triadic closure (if student A interacts with students B and C, then an interaction between students B and C forms triadic closure)
    \item \textit{Homophily on lab section}: measure of edges occurring between students in the same lab section
    \item \textit{Homophily on discussion section}: measure of edges occurring between students in the same discussion section
    \item \textit{Main effect of final course grade on degree}: measure of individuals' total number of adjacent edges as related to their final course grade 
    \item \textit{Homophily on gender}: measure of edges occurring between students of the same gender
    \item \textit{Main effect of gender on degree (woman)}: measure comparing women's total number of adjacent edges to men's total number of adjacent edges
    \item\textit{Homophily on race/ethnicity}: measure of edges occurring between students of the same URM status
    \item \textit{Main effect of race/ethnicity on degree (URM)}: measure comparing URM students' total number of adjacent edges to non-URM students' total number of adjacent edges
\end{enumerate}

For each of the eight observed networks, we determined the coefficient estimates of these ten predictor variables using MCMC MLE in the \textit{ergm} package in R. We describe how we determined the model's goodness-of-fit in the Appendix. 

We note that, particularly in EM-Phys, some sample sizes (especially across gender and racial/ethnic groups) seem too small to make statistical comparisons. ERGMs, however, consider edges and not nodes as the unit of analysis. Although the number of nodes is small, the network is quite dense and includes many of the possible edges. Smaller sample sizes, furthermore, do not prevent valid estimation of the coefficient values. Rather, they are reflected in the standard errors and $p$-values of the coefficients~\cite{kolaczyk2015question}. Quantitative modifications to ERGMs are only necessary for very small networks (less than six nodes)~\cite{yon2021exponential}.

\vspace{-0.75pc}
\section{Results}
\vspace{-0.5pc}

In this section, we first describe the densities and structures of the networks in each instructional context. We then present statistical results about whether and how students' lab and discussion section enrollment, final course grades, gender, and race/ethnicity relate to the formation of the observed networks.



\vspace{-0.3cm}
\subsection{Structural comparisons}
\vspace{-0.3cm}


\begin{figure}[t]
    \centering
    \includegraphics[width=3in]{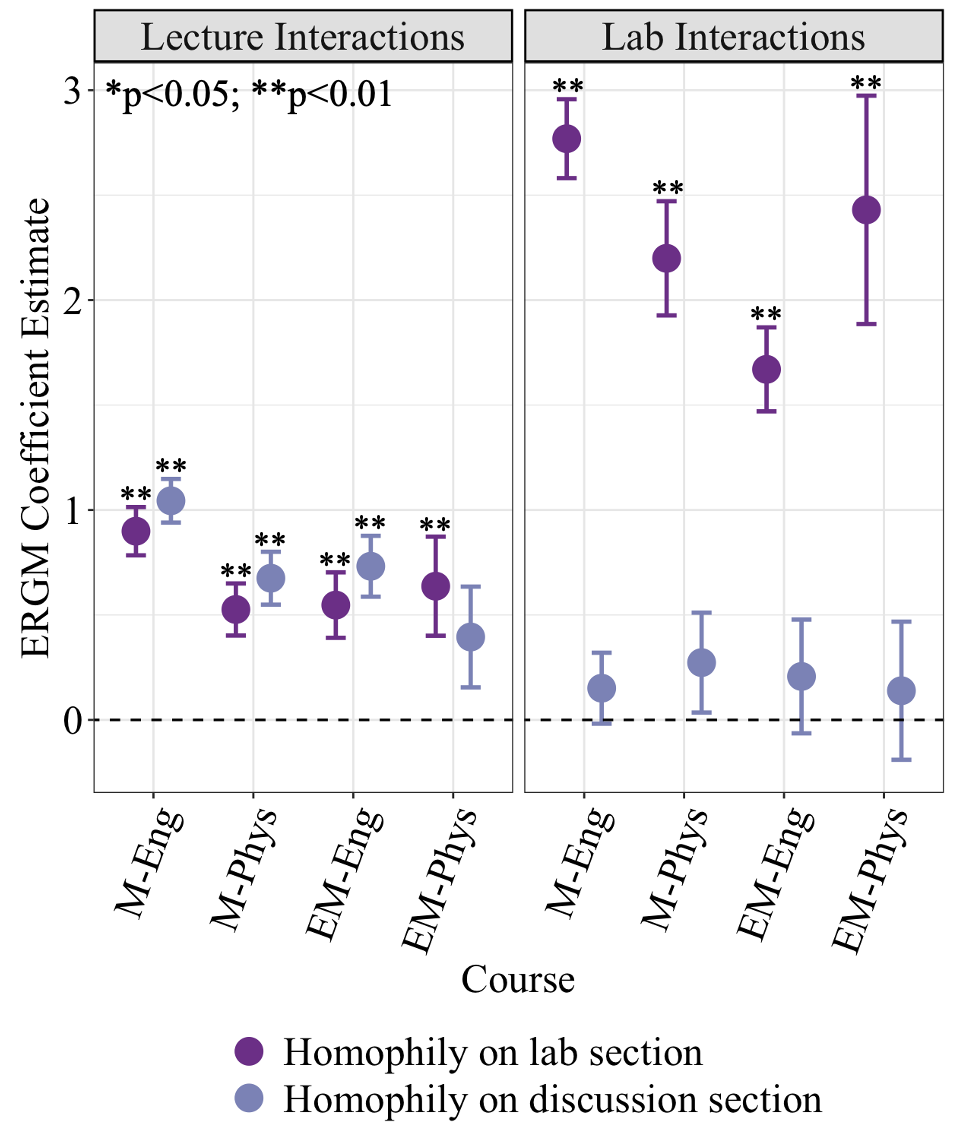}
    \caption{Plot of ERGM coefficients for the \textit{homophily on lab section} and \textit{homophily on discussion section} variables. A more positive (negative) coefficient estimate indicates that more (fewer) edges occur between students in the same section. Error bars indicate the standard error for each estimate and asterisks indicate statistical significance.}
    \label{fig:sectionplot}
\end{figure}

Within each course, the densities of the lecture and lab interaction networks (listed in Figs. \ref{fig:fallsociograms} and \ref{fig:springsociograms}) are comparable to an order of magnitude. These densities indicate a roughly similar level of connectedness among students in both instructional contexts, however the structure of these connections varies. The lecture interaction networks (shown in the left-hand column of Figs. \ref{fig:fallsociograms} and \ref{fig:springsociograms}) contain long chain-like formations that connect many nodes in one large cluster, with some additional, smaller clusters not connected to this main cluster. This structure is consistent across courses using active learning techniques (M-Eng and EM-Eng) and traditional instruction methods (M-Phys and EM-Phys) in lectures. In contrast, the lab interaction networks (shown in the right-hand column of Figs. \ref{fig:fallsociograms} and \ref{fig:springsociograms}), contain smaller chain-like formations and many disconnected clusters of two to four nodes, likely indicating separation by lab group. 



\begin{figure}[t]
    \centering
    \includegraphics[width=3.3in]{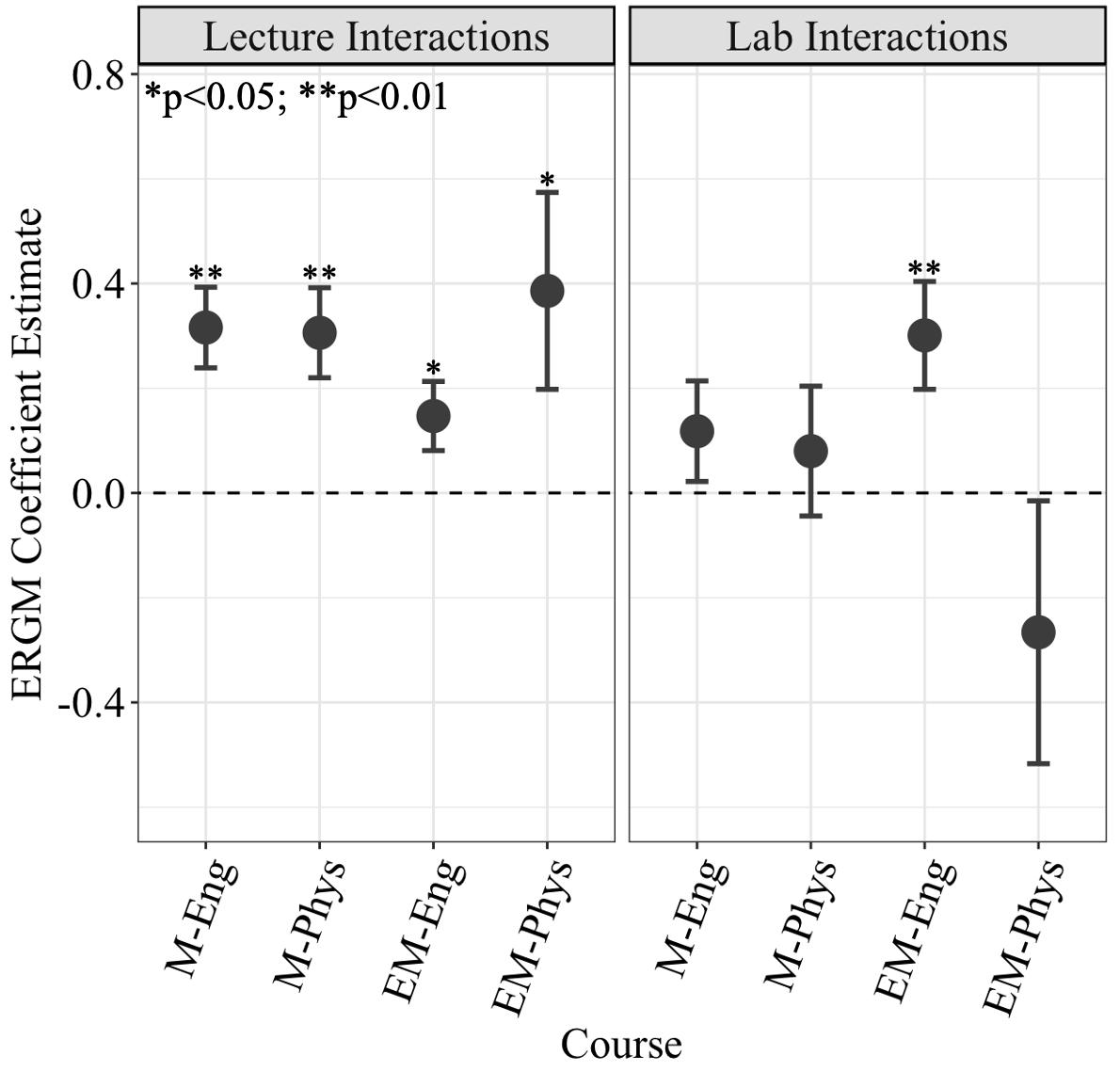}
    \caption{Plot of ERGM coefficients for the \textit{main effect of final course grade on degree} variable for each observed network. A more positive (negative) coefficient estimate indicates that students with higher final course grades have more (fewer) total connections in the network than students with lower final course grades. Error bars indicate the standard error for each estimate and asterisks indicate statistical significance.}
    \label{fig:gradeplot}
\end{figure}

Coefficient estimates for the first three variables in our ERGM expand on these visual interpretations (Appendix Table~\ref{tab:coefficients}). The coefficient estimates for the \textit{edges} variable (or main intercept) indicate that there are significantly fewer edges present in every observed network than we would expect if the edges were formed randomly. It is typical of most social networks to have fewer edges than expected at random~\cite{toivonen2006model}. In addition, with the exception of the M-Phys lab interaction network, all networks contain a significant number of reciprocal edges (\textit{reciprocity} variable), meaning that pairs of students frequently report interacting with each other. There is also a strong presence of triadic closure (\textit{GWESP} variable) in all networks except the EM-Phys lab interaction network, suggesting group-like structures or connections among small subsets of students. Based on these measures of reciprocity and triadic closure, therefore, collaboration between groups of two or three students is characteristic of both lecture and lab interaction networks. As described above, however, whether these smaller groups are chained together in larger clusters (lecture) or remain isolated (lab) is what distinguishes the network structures of the two instructional contexts.


\vspace{-0.75pc}
\subsection{Lecture interaction network formation}
\vspace{-0.5pc}

Interactions about lecture material frequently occur between peers in the same lab and discussion sections (left panel of Fig. \ref{fig:sectionplot}). In all four courses, students discuss lecture material with peers enrolled in their lab section significantly more than with peers not in their lab section (dark purple dots on the left panel of Fig. \ref{fig:sectionplot}). Additionally, with the exception of EM-Phys, students have a strong tendency to discuss lecture material with peers in their  discussion section (light purple dots on the left panel of Fig. \ref{fig:sectionplot}). 

\begin{figure}[t]
    \centering
    \includegraphics[width=3.1in]{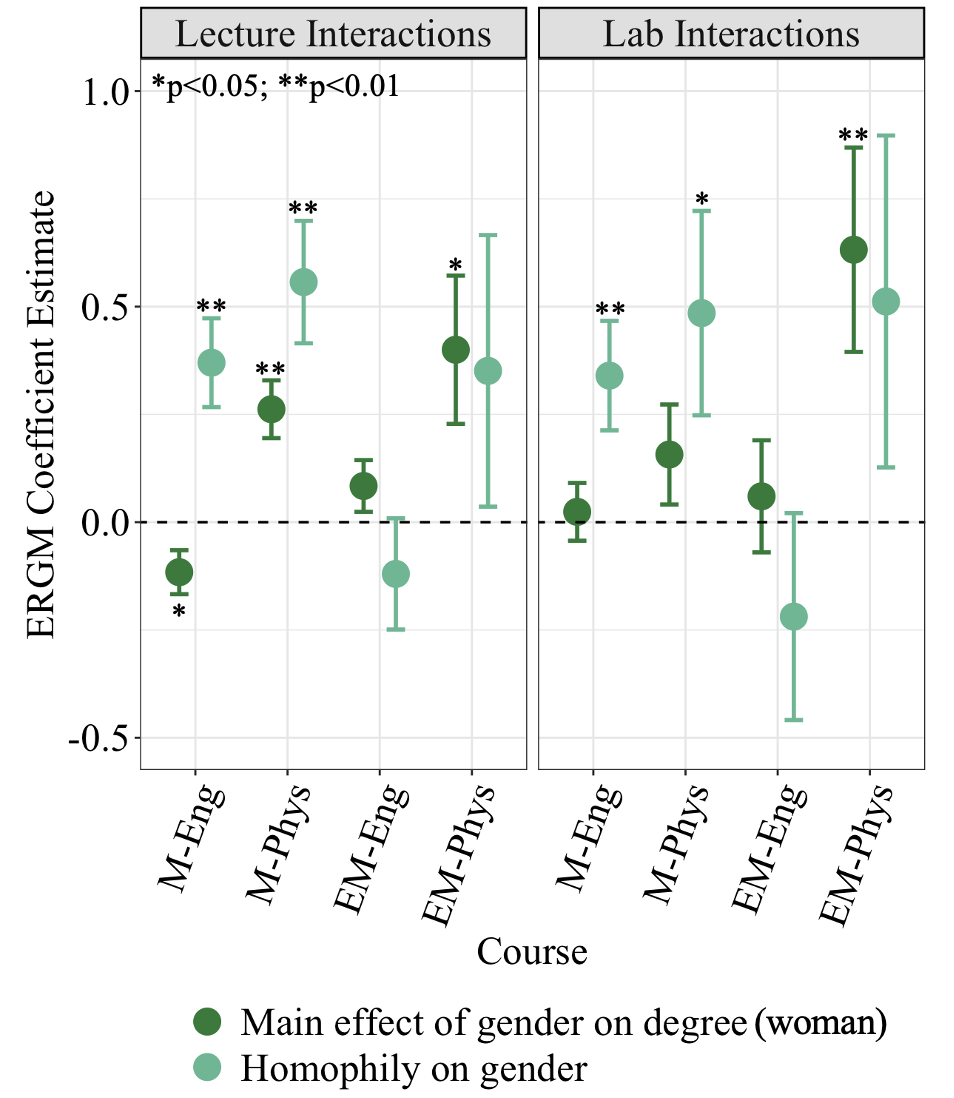}
    \caption{Plot of ERGM coefficients for the two predictor variables related to gender. A more positive (negative) coefficient estimate for the \textit{main effect of gender on degree} variable indicates that women have more (fewer) connections than men. A more positive (negative) coefficient estimate for the \textit{homophily on gender} variable indicates that more (fewer) edges occur between students of the same gender. Error bars indicate the standard error for each estimate and asterisks indicate statistical significance.}
    \label{fig:genderplot}
\end{figure}

\begin{figure}[t]
    \centering
    \includegraphics[width=3.1in]{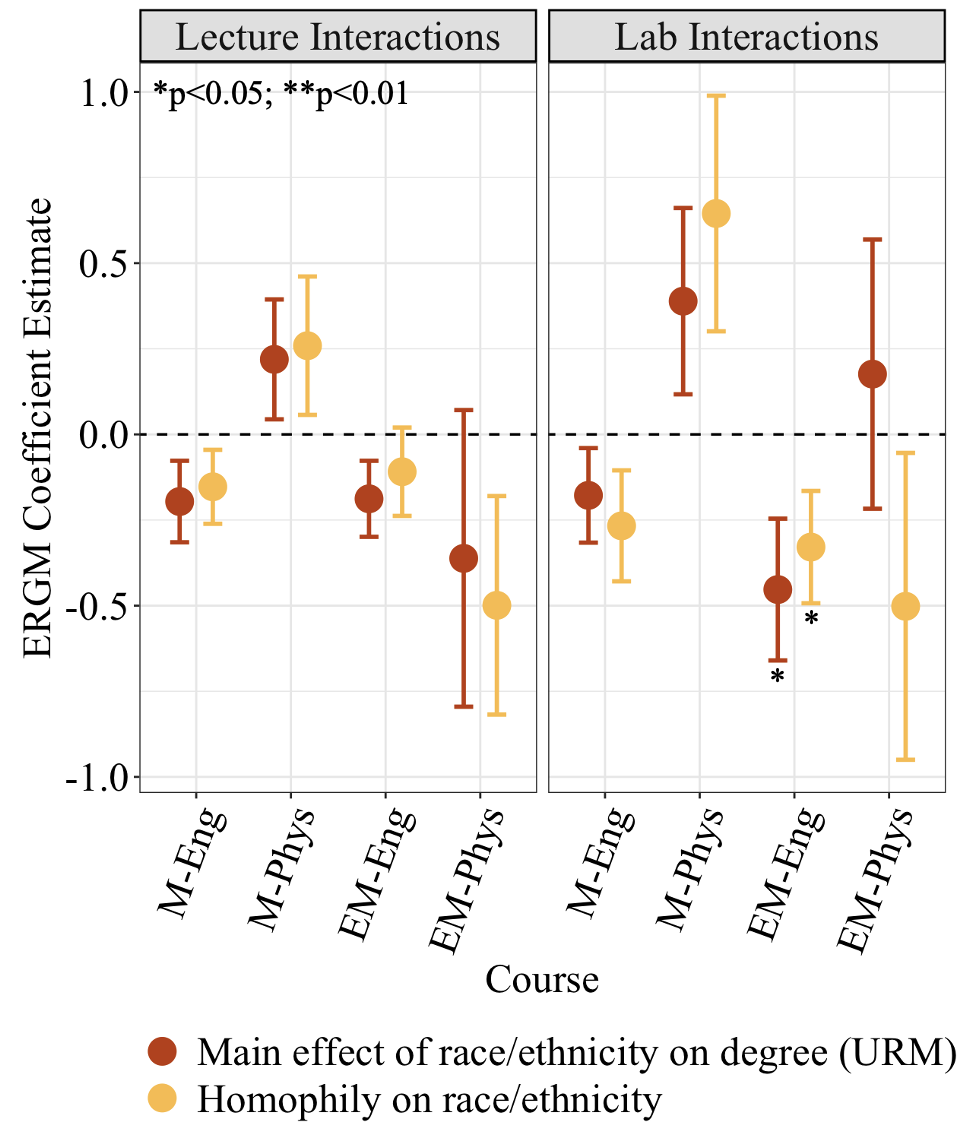}
    \caption{Plot of ERGM coefficients for the two predictor variables related to race/ethnicity. A more positive (negative) coefficient estimate for the \textit{main effect of race/ethnicity on degree} variable indicates that URM students have more (fewer) connections than non-URM students. A more positive (negative) coefficient estimate for the \textit{homophily on race/ethnicity} variable indicates that more (fewer) edges occur between students of the same race/ethnicity. Error bars indicate the standard error for each estimate and asterisks indicate statistical significance.}
    \label{fig:urmplot}
\end{figure}

Lecture interactions are also more frequent for high-achieving students (left panel of Fig. \ref{fig:gradeplot}). In all four courses, students with higher final course grades tend to have more interactions about lecture material than students with lower final course grades. 

Gender is related to lecture interaction networks in two ways: whether students of different genders have different numbers of connections and whether students tend to interact with peers of their same gender (left panel of Fig. \ref{fig:genderplot}). First, in M-Eng, women have significantly fewer lecture connections than men (dark green dots on the left panel of Fig. \ref{fig:genderplot}). In EM-Eng, women and men have comparable numbers of lecture connections. In M-Phys and EM-Phys, women have significantly more lecture connections than men. Second, students in M-Eng and M-Phys tend to interact with peers of their same gender, however men and women in EM-Eng and EM-Phys proportionately interact with one another (light green dots on the left panel of Fig. \ref{fig:genderplot}).

Finally, students do not interact with peers about lecture material on the basis of race/ethnicity. In all four courses, URM and non-URM students have comparable numbers of lecture connections (red dots on the left panel of Fig. \ref{fig:urmplot}) and proportionately interact with one another (yellow dots on the left panel of Fig. \ref{fig:urmplot}).

\vspace{-0.75pc}
\subsection{Lab interaction network formation}
\vspace{-0.5pc}

Lab section enrollment, but not discussion section enrollment, relates to the formation of lab interaction networks (right panel of Fig. \ref{fig:sectionplot}). In all four courses, students have a significant tendency to interact with peers in their lab section about lab material (dark purple dots on the right panel of Fig. \ref{fig:sectionplot}). Students in every course, however, proportionately interact with peers in and not in their discussion section about lab material (light purple dots on the right panel of Fig. \ref{fig:sectionplot}). These patterns are distinct from the lecture interaction networks, where students frequently discuss lecture material with peers in both their lab and discussion sections.

Students' final course grades are mostly unrelated to their position in the lab interaction networks. In M-Eng, M-Phys, and EM-Phys, students of all levels of achievement (e.g., low or high final course grades) have comparable numbers of lab connections (right panel of Fig. \ref{fig:gradeplot}). In EM-Eng, however, high-achieving students have significantly more lab connections than their low-achieving peers. With this one exception, final course grades offer another difference between how interaction networks form in each context: students with higher final course grades tend to have more lecture connections, but they do not necessarily have more lab connections.

Gender is also related to the formation of lab interaction networks. In M-Eng, M-Phys, and EM-Eng, men and women have comparable numbers of lab connections (dark green dots on the right panel of Fig. \ref{fig:genderplot}). In EM-Phys, women have significantly more lab connections than men. Network connections, therefore, are more equally distributed between men and women in the lab context than in the lecture context. In addition, students in M-Eng and M-Phys have a significant tendency to interact with peers of their same gender about lab material (light green dots on the right panel of Fig. \ref{fig:genderplot}). Men and women in EM-Eng and EM-Phys, however, proportionately interact with one another about lab material. This trend is the same as in the lecture interaction networks.

Lastly, students mostly do not interact with peers about lab material on the basis of race/ethnicity. In M-Eng, M-Phys, and EM-Phys, URM and non-URM students have similar numbers of lab connections (red dots on the right panel of Fig. \ref{fig:urmplot}) and proportionately interact with one another (yellow dots on the right panel of Fig. \ref{fig:urmplot}). In EM-Eng, however, URM students have significantly fewer lab connections than their non-URM peers and students tend to interact with peers of a different race/ethnicity about lab material. With this one exception, the relationship between students' race/ethnicity and network formation is the same in both the lecture and lab instructional contexts.

\vspace{-0.3cm}
\section{Discussion}
\vspace{-0.3cm}

In this study, we used ERGMs to determine whether and how students' section enrollment, final course grade, and demographics relate to the formation of interaction networks in four different remote physics courses. We found that these variables had different relationships to network formation in lecture and lab instructional contexts, offering multiple explanations for the different network structures we observed. 

\vspace{-0.75pc}
\subsection{Network structure}
\vspace{-0.5pc}

We found that the lecture and lab interaction networks contained similar levels of connectedness, however the structure of these connections varied. Specifically, the lecture interaction networks centered around one large cluster of students connected along chains of edges. The lab interaction networks, however, contained many small and disconnected clusters of students. Interestingly, these distinct structures of lecture and lab interaction networks arise in both in-person~\cite{commeford2021characterizing} and remote (our study) physics courses. Patterns of interactions about lecture and lab material, therefore, seem to form differently, regardless of the modality.

Prior work suggests that the instructional style of lectures, whether active or traditional, might explain the different network structures~\cite{brewe2010changing,commeford2021characterizing,traxler2020network}. In our study, however, we observed similar structures in all four lecture interaction networks despite two of the lectures employing active learning techniques and the other two lectures following traditional instruction. This contradiction agrees with others~\cite{commeford2021characterizing} who argue that, while instructional style may relate to network structure, broad categories of instructional styles (e.g., active and traditional) do not offer a sufficient explanation~\cite{commeford2022latent,stains2018anatomy}. Instead, it is necessary to examine more fine-grained instructional differences or student-level variables, which we elaborate on below. 



\vspace{-0.75pc}
\subsection{Section enrollment}
\vspace{-0.5pc}

Beyond instructional style, the nature of student interactions varies between the different components of a course. We found that students discussed lecture material with peers in both their lab and discussion sections, but they only discussed lab material with peers in their lab section. One explanation for these results might be that students' lab and discussion peers were the same people. If that were the case, however, we would observe students discussing both lecture and lab material with both lab and discussion peers, which we did not.

Instead, we propose two explanations for the different relationships between section enrollment and the lecture and lab interaction networks: the distinct learning goals and levels of pervasiveness of each instructional context. First, the learning goals for lectures and discussion sections in all four courses were for students to understand physics concepts and solve problems about these concepts. The learning goals for the labs, on the other hand, aimed to develop students' experimental skills and scientific decision-making and explicitly did not reinforce lecture content (as per, e.g., ~\cite{Phys21,kozminski2014aapt,holmes2018introductory,Smith2021}). It is sensible, therefore, that we observed students discussing the material relevant to each context with peers in the corresponding class sections (e.g., interacting with lab peers about lab material and with discussion peers about lecture material). The distinct learning goals also explain why interactions about lab material did not take place during discussion sections or with discussion peers: lab content was not relevant to discussion work.



We speculate that the pervasiveness of lecture material explains why students discussed lecture material with lab peers. 
Outside of lectures and discussion sections, students completed weekly pre-reading quizzes and written homework assignments, studied for exams, and attended office hours for help. The lecture material, therefore, remained salient at all times, including during lab section when students could interact with other peers in the course. Students also had more time during labs to have conversations about the broader course because lab sections were two hours long (lectures and discussion sections were 50 minutes long). In contrast, labs were not as pervasive because the experiments and lab notes were completed during lab sections and the individual homework assignments were short. This likely reduced students' need to interact with peers about lab material outside of class time. Future work should probe this explanation more directly, such as by asking students to comment on the nature of their interactions with their peers on the network survey.


\vspace{-0.75pc}
\subsection{Final course grades}
\vspace{-0.5pc}

Previous studies have unanimously shown that students holding a more central position in an interaction network, whether course-wide or within labs, achieve higher learning outcomes~\cite{williams2019linking, williams2015understanding,dokuka2020academic,grunspan2014,bruun2013talking,traxler2018networks,stadtfeld2019integration}. We might expect, therefore, that lecture interactions correlate with students' learning of lecture material and that lab interactions correlate with student's learning of lab material. In our study, we found that students with more lecture connections tended to have higher final course grades than their peers with fewer lecture connections, replicating previous work on in-person physics courses. Students with more lab connections, however, did not systematically earn higher final course grades than their peers with fewer lab connections. 

We surmise that these results are due to the relative weightings of lecture and lab material in students' final course grades. In all four courses, lecture material (exams, homework, participation in discussions, etc.) accounted for at least 80\% of students' final course grades. It is unsurprising, therefore, that students with more lecture interactions also earned higher final course grades: if students' interactions about lecture material improved their learning of lecture concepts, then that learning would be largely captured by these grades. We note that our statistical models do not infer causal relationships (i.e., having more lecture interactions might not necessarily cause the learning of lecture concepts); rather, our proposed explanation is conjectural. Lab assignments (individual homework and group lab notes), on the other hand, only accounted for between 10\% and 20\% of students' final course grades in each course. Additionally, the lab notes were graded at the group level: all students within a group received the same grade regardless of their individual contributions. With this low weighting and group-level grading, even if interacting about lab material with peers helped individual students' performance in lab as prior work suggests~\cite{wei2018developing,park2017chemical,blickenstaff2010framework}, our statistical analysis likely did not catch it. 



There was one anomaly in these results, namely that number of lab connections was positively correlated with final course grade in EM-Eng. One explanation for this is that interacting with peers about lab material helped students master lecture material in this course. Prior work, however, suggests that this explanation is unlikely because the two contexts have distinct learning goals~\cite{etkina2007,smith2020direct,holmes2017added,adams2015analyzing}. Our results for EM-Phys also refute this explanation: if lab interactions helped students learn material from electromagnetism lectures then we would observe a similar result in EM-Eng, which we do not. Alternatively, this observation could be a statistical signal that students with more lab interactions earned higher lab grades. EM-Eng weighted labs more than M-Eng and M-Phys (20\% versus 10\%), in which we observed no significant correlation between number of lab connections and final course grade. We might have resolved a relationship between lab interactions and lab performance in EM-Eng, therefore, because labs were given more weight in the final course grades. This correlation would agree with previous studies suggesting that interacting with peers about lab material improves students' performance in labs~\cite{wei2018developing,park2017chemical}. EM-Phys, however, also weighted labs as 20\% of the final course grades and we observed no correlation between number of lab connections and final course grade in that course. While the low survey response rate for this course (about 60\%) leaves this result only tentative, this finding provides evidence against our second explanation. Future work should further investigate the relationship between the content of students' interactions and students' academic performance across instructional contexts. 


\vspace{-0.75pc}
\subsection{Gender}
\vspace{-0.5pc}

Previous studies found conflicting results related to whether men and women have different or comparable numbers of connections in course-wide interaction networks~\cite{dokuka2020academic,williams2015understanding,brewe2012investigating,dou2016beyond,wells2019}. Our study adds nuance to these findings, suggesting that in both remote and in-person courses such gender-based patterns likely depend on the student population of a course -- the composition of enrolled students' genders and majors -- and/or the structure of assignments -- whether submitted in groups or individually. 

The gender balance and, relatedly, majors of students enrolled in a course seem related to the formation of lecture (and in other studies, course-wide) interaction networks. In gender-balanced or majority-women courses (M-Eng and EM-Eng in our study and the in-person courses in Ref.~\cite{dokuka2020academic,williams2015understanding,brewe2012investigating,dou2016beyond}), either men and women had comparable numbers of lecture connections or men had more lecture connections than women. When a minority of students in a course are women (M-Phys and EM-Phys in our study and, presumably, the in-person course in Ref.~\cite{wells2019}), however, women had more lecture connections than men. Importantly, the gender composition of science courses also tends to correlate with students' majors, therefore we cannot disentangle these two explanations. In our study, for example, students in the courses for physics majors (M-Phys and EM-Phys) were majority men and the courses for non-physics majors (M-Eng and EM-Eng) contained gender-balanced enrollment. These results indicate that students in the minority gender group of a class often engage in more interactions than their peers and that this phenomenon is typical of science courses intended for students majoring in the discipline. Future work should examine whether and how peer interactions support the learning experiences of such underrepresented and minoritized students.



Additionally, our results pertaining to gender suggest that the structure of assignments within an instructional context might relate to network formation. In three out of four lab interaction networks and in the EM-Eng lecture interaction network, men and women proportionately engaged in peer interactions. In each of these contexts, students completed and submitted assignments in small groups. We found in the remaining three lecture interaction networks, however, that men and women had significantly different numbers of network connections. In these contexts, students completed work in small groups but submitted them individually.  We note that the EM-Phys labs depended on group work and that women had more connections than men in this network, offering a possible contradiction. This result is only preliminary because of the low survey response rate for this course (about 60\%). Future work should investigate whether and how assignment structure is also related to network formation in in-person courses.



We also found in both the lecture and lab interaction networks that students tended to interact with peers of their same gender in the mechanics courses (M-Eng and M-Phys), but not in the subsequent electromagnetism courses (EM-Eng and EM-Phys). We speculate that students in the mechanics courses of our study, most of whom were entering their first semester of college, interacted based on the guiding principle of homophily~\cite{mcpherson2001birds}. This principle contends that interactions between people of similar attributes (e.g., gender) are more common than interactions between people with different attributes. Indeed, another study found gender homophily within the interaction networks of first-year students in in-person economics courses~\cite{dokuka2020academic}. We observed in the subsequent electromagnetism course the following semester, however, that students no longer tended to interact with peers of the same gender. Given that many students in the observed electromagnetism courses also took one of the observed mechanics courses, we infer that this increase in interactions between students of different genders could be due to social integration. That is, students likely became familiar with diverse peers during the mechanics courses. 

\vspace{-0.75pc}
\subsection{Race/ethnicity}
\vspace{-0.5pc}

We generally found that students' race/ethnicity was not related to interaction network formation. In seven out of eight observed networks, URM and non-URM students had comparable numbers of connections and proportionately interacted with one another. This observation may be surprising given that URM students have been found to have a lower sense of belonging than their non-URM peers during remote instruction~\cite{conrad2021}. Previous studies~\cite{zwolak2017students,williams2015understanding,brewe2012investigating}, however, found that students' race/ethnicity was not a significant predictor of their network position in in-person courses when the majority of students were URM. Our study, therefore, provides evidence that students' race/ethnicity still does not strongly relate to their patterns of interaction during remote instruction when most students are non-URM.

The one exception to our claim above was the lab interaction network in EM-Eng. In this network, URM students had significantly fewer connections than their non-URM peers and there was a strong tendency for students to interact with peers of a different race/ethnicity. This might add nuance to our claim above that race/ethnicity does not relate to the formation of interaction networks, namely that URM students may be marginalized in lab interactions. This relationship was not observed in the other three lab networks, however, leaving possible claims only tentative -- this result may simply be a statistical fluctuation.

\vspace{-0.75pc}
\subsection{Limitations}
\vspace{-0.5pc}

We conclude this section by acknowledging a few limitations to the study. First, all data reported here were collected during a global pandemic. Students' emotional lives, learning experiences, and interactions with peers were strongly impacted by this pandemic and the transition to remote instruction~\cite{wilcox2020recommendations,hussein2020exploring,karalis2020teaching,kyne2020covid,gillis2020covid19,dew2021student,klein2021studying,doucette2021newtothis,marzoli2021,conrad2021,rosen2021}. These circumstances might have influenced the extent to which students engaged in, recalled, and reported meaningful interactions with peers on the survey we administered. We found similar network structures to those observed in in-person courses~\cite{commeford2021characterizing}, however, suggesting commonalities between students' interaction patterns in both settings.

The network survey itself carried other limitations. Students completed the survey as part of a homework assignment. While other studies have similarly asked students to complete network surveys online and outside of class~\cite{commeford2021characterizing,traxler2020network}, students may have reported different interactions if they had completed the survey in class when surrounded by their peers. In addition, when administering the survey, we did not provide the names of students in the course to respondents. Students may have recalled meaningful interactions with peers but could not remember the names of these peers. There could also be recall bias, where students failed to remember and report meaningful interactions. These factors may have led to under-reporting of meaningful interactions. The nature of the online instruction over Zoom, however, meant students readily had access to peers' names -- more readily than in-person instruction. Finally, we relied on surveys administered at one point in time. Networks evolve over the course of a semester \cite{brewe2010changing,grunspan2016}, so while our analysis captures one snapshot of these networks in detail, they may have changed by the end of the courses. Future work may collect and analyze similar surveys at more points in time during a semester to examine the network dynamics.

Lastly, this study was conducted at a private, research-based institution and offers a glimpse of the nature of interactions with that population and the types of instruction employed there. Future work should continue to probe additional student populations and instructional contexts. The results here already build on previous work with other institutions and types of instruction, demonstrating potential commonalities and differences between these contexts.

\vspace{-0.75pc}
\section{Conclusion}
\vspace{-0.5pc}

Previous research on in-person physics courses suggests that differences in interaction network structure between the instructional contexts of lecture and lab may be related to differences in instructional style and/or a handful of student-level variables. We investigated these possible relationships further by examining lecture and lab interaction networks in four different remote physics courses serving various student populations. We also applied statistical analysis methods, exponential random graph models, that have recently emerged in the PER community but offer a promising avenue for further work. 

We observed very similar network structures to prior studies of in-person physics courses. Results suggest that these network structures likely relate to a combination of variables: the learning goals of various instructional contexts, the pervasiveness of different course material, students' grades, whether assignments are completed in groups or individually, the distribution of gender and major of students enrolled in a course, and the tendency for students to interact with peers of their same gender. Interestingly, our study agrees with prior work that students' race/ethnicity does not correlate with their position in an interaction network. 

We are currently collecting more network surveys with an additional question asking students to write the specific content about which they interacted with their peers. We will use these responses to characterize the content of student interactions within each instructional context. We will then form multi-layer interaction networks, for example distinguishing interactions about homework from those about exam preparation. We plan to relate these different interactions to student performance to determine whether certain kinds of interactions are more central to learning, as called for by recent work~\cite{traxler2022networks}. We will conduct the research in the same course sequences as this study, but now with the lab as a standalone course. This new structure will allow us to disentangle the relationship between lecture and lab interactions and student performance in those contexts.

\vspace{-0.75pc}
\section*{ACKNOWLEDGEMENTS}
\vspace{-0.5pc}
This material is based upon work supported by the National Science Foundation Graduate Research Fellowship Program Grant No. DGE-2139899 and Grant No. DUE-1836617. We thank Cole Walsh, Barum Park, Matthew Dew, Eric Brewe, and David Esparza for engaging in meaningful discussions about this work.

\bibliography{InteractionsBibliography.bib} 

\clearpage

\section{Appendix}

\vspace{-0.75pc}
\subsection{Survey text processing}
\vspace{-0.5pc}

The interaction survey asked students to list the names of all peers with whom they interacted about the course in one text box. Some students just wrote their peers' names, while others wrote sentences with names embedded in the text. To extract the names of students listed, we took each response and split it on common delimiters (periods, commas, spaces, line breaks, semicolons) to get a list of tokens in the response (e.g., ``the cat is red" turns into [``the", ``cat", ``is", ``red"]) and compared each of the tokens to each of the first and last names from the class roster. We connected the token to a student on the class roster only if there was an exact match. We then checked pairs of tokens for matches on full name (e.g., pairs of tokens from the last example are [``thecat", ``catis", ``isred"]), matching up names for which fewer than 0.3 times the length of someone's full name corrections were needed to match the full name on the roster. We chose the constant 0.3 via trial and error, finding that this worked best for capturing as many close matches as possible without producing false negatives. If a name (either first or last) appeared multiple times in the data set, then we did not match on listings of just that name itself and we only matched listings of the other half of the name or the full name.

\vspace{-0.75pc}
\subsection{ERGMs: Goodness of fit}
\vspace{-0.5pc}

\begin{figure}[t]
    \centering
    \includegraphics[scale=0.6]{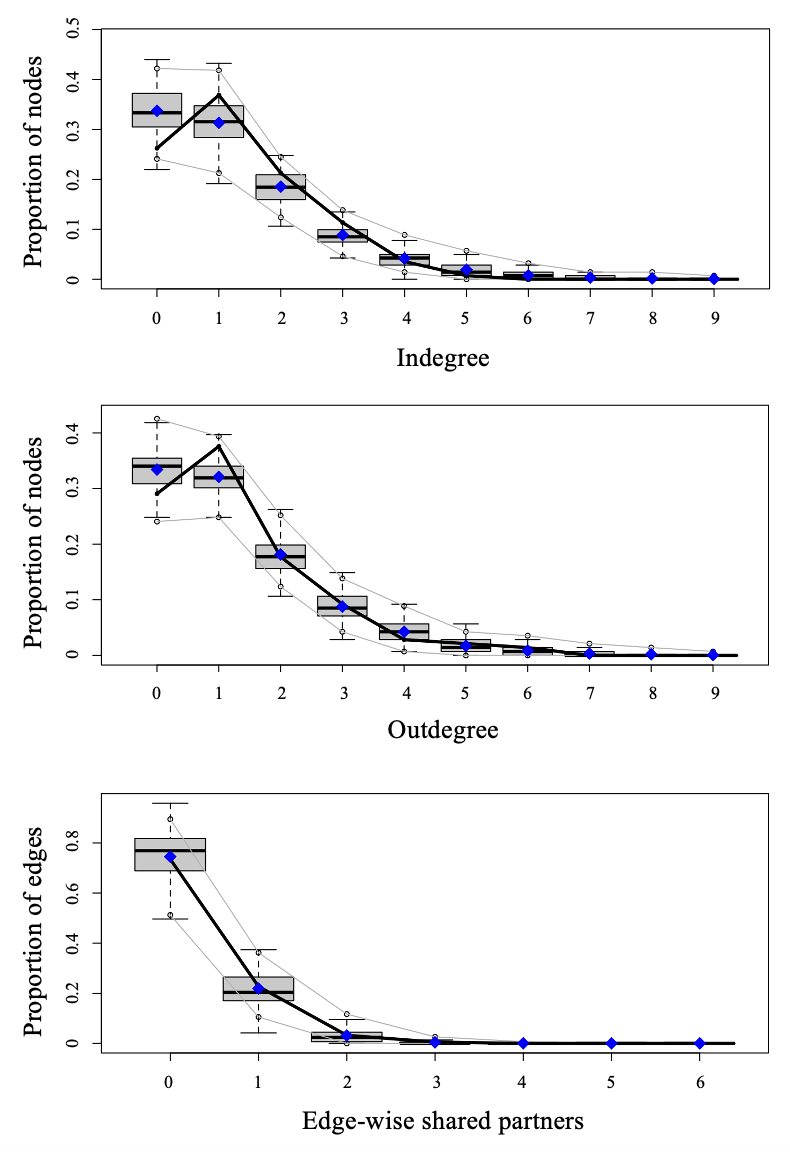}
    \caption{Goodness-of-fit plots for the M-Eng lab interaction network. The horizontal axis represents the value of the network measure and the vertical axis represents frequency. Plots compare the distribution of each measure for students in the observed data (thick black line) to that for 10 network simulations generated using the coefficient estimates of the model (boxplots).}
    \label{fig:gofplot}
\end{figure}

To determine whether an ERGM fits an observed network well, we compare our observed network to random networks simulated using the model's predictor variables and coefficient estimates. For example, Fig. \ref{fig:gofplot} shows the distributions of three different network measures – indegree, outdegree, and edge-wise shared partners (measure of edges forming triadic closure) – for the observed M-Eng lab interaction network (shown as a black line) and ten simulated networks (shown as boxplots) using the model coefficients for this network. 
For all three measures, the distribution for the observed network falls within the boxplots of simulated networks, indicating a good fit of the model to the observed network. We observed similar plots for the other seven networks as well.

\vspace{-0.75pc}
\subsection{ERGM coefficient estimates}
\vspace{-0.5pc}

Table \ref{tab:coefficients} summarizes the coefficient estimates of the ERGM model for each network. We interpret the coefficient estimates as log-odds of edge formation. For example, the coefficient estimate for the \textit{homophily on lab section} variable for the M-Eng lecture interaction network is 0.90. This means that the log-odds of an edge forming in the network increases by 0.90 for each additional edge connecting students in the same lab section, holding the rest of the network the same. In other words, edges connecting students in the same lab section are more probable than edges connecting students in different lab sections, even after accounting for the other configurations included in the model.

\begin{table*}[t]
\caption{\label{tab:coefficients}%
Coefficient estimates for our ERGM fit to the four observed courses (eight observed networks). Standard errors of the coefficient estimates are in parentheses below. Asterisks indicate statistical significance ($^{*}$p$<$0.05; $^{**}$p$<$0.01).}
\setlength{\extrarowheight}{1pt}
\setlength{\tabcolsep}{5pt}
\begin{tabular}{lcccccccc}
\hline
\hline \\[-1.8ex] 
 & \multicolumn{2}{c}{M-Eng} & \multicolumn{2}{c}{M-Phys} & \multicolumn{2}{c}{EM-Eng} & \multicolumn{2}{c}{EM-Phys}
\\
Variable & Lecture & Lab & Lecture & Lab & Lecture & Lab & Lecture & Lab\\
\hline
\textit{Edges} & -8.13$^{**}$ & -7.39$^{**}$ & -7.66$^{**}$ & -7.35$^{**}$ & -6.71$^{**}$ & -7.67$^{**}$ & -7.22$^{**}$ & -4.41$^{*}$\\
 & (0.59) & (0.73) & (0.70) & (0.99) & (0.47) & (0.75) & (1.38) & (1.80)\\
 \\
\textit{Reciprocity} & 3.54$^{**}$ & 2.42$^{**}$ & 3.40$^{**}$ & 1.03 & 3.85$^{**}$ & 2.94$^{**}$ & 3.16$^{**}$ & 2.94$^{**}$\\
 & (0.26) & (0.37) & (0.31) & (0.56) & (0.21) & (0.40) & (0.52) & (0.65)\\
 \\
\textit{GWESP (triadic closure)} & 1.31$^{**}$ &  1.12$^{**}$ & 0.72$^{**}$ & 1.32$^{**}$ & 1.41$^{**}$ & 1.28$^{**}$ & 0.49$^{*}$ & 0.38\\
 & (0.10) & (0.15) & (0.10) & (0.19) & (0.10) & (0.15) & (0.22) & (0.27)\\
 \\
\textit{Homophily on lab section} & 0.90$^{**}$ &  2.77$^{**}$ & 0.53$^{**}$ & 2.20$^{**}$ & 0.55$^{**}$ & 1.67$^{**}$ & 0.64$^{**}$ & 2.43$^{**}$\\
 & (0.12) & (0.19) & (0.12) & (0.27) & (0.16) & (0.20) & (0.24) & (0.54)\\
 \\
\textit{Homophily on discussion section} & 1.04$^{**}$ &  0.15 & 0.68$^{**}$ & 0.27 & 0.73$^{**}$ & 0.21 & 0.40 & 0.14\\
 & (0.10) & (0.17) & (0.13) & (0.24) & (0.15) & (0.27) & (0.24) & (0.33)\\
 \\
\textit{Main effect of final course grade on degree} & 0.32$^{**}$ & 0.12 & 0.31$^{**}$ & 0.08 & 0.15$^{*}$ & 0.30$^{**}$ & 0.39$^{*}$ & -0.27\\
 & (0.08) & (0.10) & (0.09) & (0.12) & (0.07) & (0.10) & (0.19) & (0.25)\\
 \\
\textit{Main effect of gender on degree (woman)} & -0.12$^{*}$ & 0.02 & 0.26$^{**}$ & 0.16 & 0.08 & 0.06 & 0.40$^{*}$ & 0.63$^{**}$\\
 & (0.05) & (0.07) & (0.07) & (0.12) & (0.06) & (0.13) & (0.17) & (0.24)\\
 \\
\textit{Homophily on gender} & 0.37$^{**}$ & 0.34$^{**}$ & 0.56$^{**}$ & 0.49$^{*}$ & -0.12 & -0.22 & 0.35 & 0.51\\
 & (0.10) & (0.13) & (0.14) & (0.24) & (0.13) & (0.24) & (0.32) & (0.39)\\
 \\
 \textit{Main effect of race/ethnicity on degree (URM)} & -0.20 &  -0.18 & 0.22 & 0.39 & -0.19 & -0.45$^{*}$ & -0.36 & 0.18\\
 & (0.12) & (0.14) & (0.18) & (0.27) & (0.11) & (0.21) & (0.43) & (0.39)\\
 \\
\textit{Homophily on race/ethnicity} & -0.15 &  -0.27 & 0.26 & 0.65 & -0.11 & -0.33$^{*}$ & -0.50 & -0.50\\
 & (0.11) & (0.16) & (0.20) & (0.34) & (0.13) & (0.16) & (0.32) & (0.45)\\
\hline
\hline \\[-1.8ex] 
\textit{}  & 
\end{tabular}
\end{table*}

\end{document}